\documentstyle[11pt,aaspp4]{article}

\def\solar{\ifmmode_{\mathord\odot}\else$_{\mathord\odot}$\fi}

\slugcomment{Accepted for Ap.J. Vol. 502, July 20, 1998}
\lefthead{Harris et al.}
\righthead{Dwarf Carbon Stars}

\begin{document}

\title{Astrometry and Photometry for Two Dwarf Carbon Stars}

\author{Hugh C. Harris, Conard C. Dahn, Richard L. Walker,
Christian B. Luginbuhl, \\
Alice K.B. Monet, Harry H. Guetter, Ronald C. Stone, Fredrick J. Vrba, \\
David G. Monet, and Jeffrey R. Pier}

\affil{U.S. Naval Observatory, Flagstaff, AZ 86002}

\begin{abstract}
Preliminary trigonometric parallaxes and BVI photometry are
presented for two dwarf carbon stars, LP765$-$18 (= LHS1075) and
LP328$-$57 (= CLS96).  The data are combined with the literature
values for a third dwarf carbon star, G77$-$61 (= LHS1555).  All
three stars have very similar luminosities
(9.6$\,<\,$M$_{\rm V}<\,$10.0) and very similar broadband colors
across the entire visual-to-near IR (BVIJHK) wavelength range.
Their visual (BVI) colors differ from all known red dwarfs,
subdwarfs, and white dwarfs.
In the M$_{\rm V}$ versus V$-$I color--magnitude diagram they are
approximately 2 magnitudes subluminous compared with normal disk
dwarfs with solar-like metallicities, occupying a region also
populated by O-rich subdwarfs with $-1.5<$[m/H]$<-1.0$.  The
kinematics indicate that they are members of the Galactic spheroid
population. The subluminosity of all three stars is due
to an as-yet-unknown combination of (undoubtedly low) metallicity,
possibly enhanced helium abundance, and unusual line-blanketing
in the bandpasses considered.
% Full analysis awaits the availablity
% of model atmospheres for metal-poor, C/O$>$1 configurations.
The properties of the stars are compared with models for
the production of dwarf carbon stars.
\end{abstract}


\section{Introduction}

The discovery of the first dwarf carbon star (\cite{dah77})
sparked considerable research on the evolution of stars
to and through the dwarf carbon star (dC) stage.  It is now
accepted that dC stars are formed through the accretion of
carbon-rich material from a close companion star that was
evolving on the asymptotic giant branch at the time of mass
transfer.   They are now recognized as members
of a larger family of stars that have undergone mass-transfer
binary evolution, a family that includes the halo CH giant stars,
the disk Ba giants, the Ba dwarf or CH subgiant stars, and
the extrinsic S stars (\cite{gre97}).

De Kool \& Green (1995) have modeled dC star formation by
following the evolution of a large variety of binary systems
constructed from the observed distributions of component
masses, orbital separations, and metallicities for unevolved
binaries. Although these simulations involve a large
number of poorly constrained parameters, the results indicate
that the formation of a dC star is strongly favored by low
initial metallicity, and that virtually no dCs are produced in systems
with metallicities above half solar.  Furthermore, the
mass distribution is found to peak strongly in the
0.4 to 0.9 solar mass range.

Despite the studies to date, our understanding of dC stars
remains highly speculative, based as much on plausibility
arguments as on objective facts.  Distances of these stars
are among the most essential missing data,  and must be known
in order to determine each star's luminosity, space velocity, mass,
and evolutionary status.
Presently, only one dC star, G77$-$61, has a reliable trigonometric
distance determination and it indicates a high space velocity
and -- by implication -- low metallicity.  This implied low metallicity
is consistent with the detailed atmospheric analysis of G77$-$61
by Gass et al. (1988) who derived the extremely low metallicity of
[Fe/H]=$-5.6$, a value which certainly implies very early
epoch halo formation.

A larger sample of dC stars with properties well constrained
by fundamental observational data is clearly needed to understand
these stars as a class.  Unfortunately, no known dC stars
are bright enough to have had their parallaxes measured with high
accuracy by the Hipparcos satellite.  In this paper we report
preliminary trigonometric parallaxes for two additional dC stars
and discuss some of their properties.

\section{Data}

The stars LP765$-$18 and LP328$-$57 have been observed since 1992
as part of the Naval Observatory CCD parallax program using a
Tektronix 2048 CCD camera on the 1.55-m Strand telescope.
This camera is producing relative
astrometry more accurate than that previously obtained using a
TI 800 CCD (\cite{mon92}), primarily as a result of the improved
reference star frames available in the larger field of view.
Experience to date demonstrates that the Tek 2048 program
generally produces relative astrometric measurements accurate
to $\pm$3 mas for a single observation and relative parallaxes
accurate to $\pm$0.5 mas after roughly 100 observations
adequately distributed over the parallactic ellipse and spanning
at least 3 years (\cite{dah97}).  This high accuracy is
essential in order to determine accurate distances for even the
nearer of the known dC stars, almost all of which are more than
100 pc from the sun.

The two dC stars reported here have 68 observations on 56 nights
over 4.2 years (LP765$-$18) and 106 observations on 86 nights
over 4.3 years (LP328$-$57).  The parallax results presented in
Table 1 are regarded as preliminary in the sense that both fields
are still being observed.  The BVI photometry reported are USNO
data taken with the 1.0-m telescope using procedures described
in Monet et al. (1992), and are accurate to $\pm\,0.02$ magnitudes.
The JHK photometric data are from Bothun et al.
(1991) and Dearborn et al. (1986).  For completeness, Table 1
includes corresponding data for G77$-$61 (the third dC star
with a measured trigonometric parallax) taken from the literature.

Also given in Table 1 is the correction ($\Delta\pi$) necessary
to convert the directly measured ``relative'' parallaxes ($\pi_{rel}$)
to ``absolute'' parallaxes ($\pi_{abs}$).  In the cases of LP765$-$18
and LP328$-$57, we have used BVI photometry of the individual
reference stars to estimate the mean parallax of each reference frame.
% which, due to the least-squares methodology in the solution for
% $\pi_{rel}$, equates directly with $\Delta\pi$.
The uncertainties in
the derived $\Delta\pi$ values were estimated as the combination of
random uncertainties ($\approx\,\sigma\,$n$^{-1/2}$, where n is the
number of stars comprising the reference frame) plus an estimate of
possible systematic errors arising from photometric and reddening 
uncertainties in each field.  Because the reference stars in
these fields typically have distances of 0.7--2 kpc, their estimated
parallaxes are small.  In the case of G77$-$61, we adopt the improved
statistical estimate for $\Delta\pi$ from van Altena et al. (1995).
As can be seen in Table 1, the estimated errors in the $\Delta\pi$
values for these stars are sufficiently small that they contribute
only a small amount to the final uncertainties in the absolute
parallaxes.

\section{Discussion}

Results interpreted from the parallax data are shown in Table 2.  In
calculating the formal uncertainties in M$_{\rm V}$ and M$_{\rm K}$
we have adopted the parallax uncertainties given in Table 1,
along with $\pm\,$0.02 mag for the uncertainties in the V magnitudes
and $\pm\,$0.03 mag for the uncertainties in the K magnitudes. The
tabulated radial velocities (V$_{\rm r}$) are from Bothun et al. (1991) for
LP765$-$18 and LP328$-$57 and from Dearborn et al. (1986) for G77$-$61.
The space velocity components (U,V,W) in Table 1 have been corrected for
a solar motion of (10, 15, 7) km s$^{-1}$ to the local standard of rest;
U is in the direction of the galactic center.

Inspection of Tables 1 and 2 reveals that these three dC stars are
remarkably similar in terms of (1) overall kinematic
properties, (2) infrared (JHK) colors, (3) optical (BVI) colors, and
(4) luminosities (M$_{\rm V}$ and M$_{\rm K}$).  Regarding the
kinematics specifically, the large overall space velocities and
the large negative V galactic velocity components are indicative
of membership in the Galactic spheroid population.

Regarding the colors, Green et al. (1992) have previously discussed
the location of dC stars in the J$-$H versus H$-$K diagram and their
apparent separation from the giant and subgiant carbon stars,
presumably due (in part) to the higher surface gravities for the dC stars.
In Figure 1 we show the location of these dC stars in the
B$-$V versus V$-$I diagram.  Because all three were originally
identified as stars with high proper motion, we include for comparison a
selection of other field stars commonly identified in proper motion
surveys: later-type degenerates (open circles), disk dwarfs (filled
circles), and metal-poor subdwarfs (crosses).  The dC stars
occupy a unique region in this diagram.  The lone exception
to a clear-cut separation for the dC stars is LP701$-$29, the only
known late-type degenerate with strong CaI absorption which blankets
the B bandpass (\cite{dah78}).  Figure 1 suggests that even broadband
photometric surveys of faint, high proper motion stars might succeed
in isolating additional dC candidates.  Their location to the
left of the most extreme metal-poor field subdwarfs known at present
suggests metallicities below [m/H]$\sim-$2.  However, quantitative
conclusions about metallicity are not possible because the field
subdwarfs have O-rich rather than C-rich atmospheres;
with the exception of the analysis of G77$-$61 by Gass et al. (1988),
no complete studies based on model atmospheres have been presented
for metal-poor, C-rich stars with dwarf-like gravities.

Figure 2 shows the location of these dC stars in the
M$_{\rm V}$ versus V$-$I color magnitude diagram.  The solid line
represents the observed mean disk main sequence as defined by USNO
parallax stars (cf. \cite{mon92}).  The dashed lines show the
metal-poor main sequences modeled by Baraffe et al. (1997) for
metallicities (scaled from solar) of [m/H]= $-1.0, -1.5$ and $-2.0.$
These authors demonstrated that their models successfully reproduce
the main sequence of several globular clusters over this range of
metallicities.  However, these models are for O-rich atmospheres
-- not the C/O$>$1 compositions which are appropriate for the dC stars.
Furthermore, the position of dC stars in color-magnitude
diagrams cannot necessarily be interpreted simply in terms of
metallicity because of the possibility of enhanced helium
abundance in their atmospheres, produced by the mass transfer event(s)
that increased their carbon abundances.  The complicated atmospheric
situation and the difficulty in uniquely determining both log$\,g$ and
the He abundance has been discussed by Gass (1988) and Gass et al.
(1988).  Therefore, until more detailed spectroscopic investigations
are undertaken to derive the helium abundance independently,
we are left with the implication from the kinematic information that low
metallicity and nearly normal helium abundances are most likely.

Figure 3 shows the location of the dC stars in the schematic M$_{\rm K}$
versus I$-$K diagram. Here the solid line represents the ``young disk''
stars defined by Leggett (1992) while the dashed lines are again the
metal-poor main sequences models from Baraffe et al. (1997).  In their
analysis of G77$-$61, Gass et al. (1988) pointed out that the J and K
fluxes should not depend strongly on the carbon abundance whereas the
fluxes in the B,V,I and H bandpasses will depend strongly on composition.
The three dC stars are also significantly subluminous with respect to
the disk main sequence in this diagram.  However, it is also clear that
going to a redder color index still does not allow a quantitative
interpretation in terms of metallicity because the location of G77$-$61
above the [m/H]=$-1.0$ curve is clearly at odds with the value of
[Fe/H]=$-5.6$ derived by Gass et al. (1988).

Figure 4 shows the location of G77$-$61 in the M$_{\rm V}$ versus 
T$_{\rm eff}$ plane using the value of T$_{\rm eff}\,=\,$4200$^{\circ}$K
derived by Gass et al. (1988).  Once again the dashed lines are from
Baraffe et al. (1997).  Here we find at least qualitative agreement
in that this extremely metal-poor object lies below the [m/H]=$-2.0$
curve.  This emphasizes the need for higher resolution spectrophotometric
observations and C-rich model atmosphere analyses for LP765$-$18 and
LP328$-$57 in order to establish (at least) T$_{\rm eff}$ values for them.

The models of Baraffe et al. (1997) indicate that the derived masses are
primarily sensitive to the absolute magnitudes (or luminosities), and only
weakly dependent on the colors.  Acknowledging the dangers of interpreting
the dC stars with O-rich models, the masses so estimated are 0.39 M\solar\ 
for LP765$-$18, 0.37 M\solar\ for LP328$-$57, and 0.30 M\solar\ for
G77$-$61.  Models for dC star formation predict that a range of masses
from 0.2 to 1.0~M{\solar} should exist (\cite{dek95}).  However, the
frequency distribution of these masses depends critically on the adopted
Initial Mass Ratio Distribution (IMRD) for the unevolved binary which is
quite uncertain.  For a flat IMRD (i.e., dN$\,\propto\,$dq, where q is
the mass ratio), the distribution for spheroid dC stars peaks rather sharply
around 0.7 M\solar\ at a space density of
dN/dlog$\,$M = 4$\,{\rm x}\,10^{-7}$ pc$^{-3}$, falling to roughly
dN/dlog$\,$M = 2$\,{\rm x}\,10^{-7}$ pc$^{-3}$ for stars in the mass range
inferred above.  On the other hand, for uncorrelated component masses,
the models predict a broader peak spanning 0.2 to 0.8 M\solar\ with
a space density of dN/dlog$\,$M = 4$\,{\rm x}\,10^{-7}$ pc$^{-3}$ for
spheroid stars.  As noted by de Kool \& Green (1995), the observed
absolute magnitudes seem to support an IMRD with uncorrelated component
masses.

However, the apparent similarity of these three dC stars
is undoubtedly influenced to some extent by selection effects.
Hotter, more massive dC stars are predicted by the models
to be common, but hotter stars will have weaker carbon bands
and may have spectra and broad-band colors that are not as
obviously different from stars with oxygen-rich atmospheres
as are these three dC stars.  Another factor that makes the
model predictions uncertain is the amount of dilution of the
carbon-rich material being transferred from the asymptotic-giant-branch
star onto its main-sequence companion:  a more massive main sequence
star (0.5--1.0~M{\solar}, for example), has a less massive
convective envelope than the stars studied here (reducing the dilution),
but the carbon-rich material may be mixed into the radiative zone
as well (\cite{pro89}), thus increasing the dilution and making
hotter dC stars more difficult to identify.

Furthermore, these three dC stars have been chosen for
parallax observations based partly on their unusually-high
proper motions.  This kinematic selection partially
accounts for their all being members of the Galactic spheroid.
Disk dC stars may be common --- the ratio of
disk/spheroid dC stars is a quantity that will help constrain
the models of dC-star formation.  Parallax data for
a sample of dC stars with smaller proper motions could help
to address this issue.  Several such dC stars are probably
close enough to get significant parallax measurements --
however, none are presently being observed in the Naval Observatory
parallax program pending time becoming available in the program
at their respective right ascensions.

We conclude that the similar properties of these three dC stars
and their membership in the Galactic spheroid may result in part
from the way these stars were selected.  Accurate distances
for a larger sample of dC stars are needed, and identification
of dC stars without kinematic bias is essential
in order to fully understand how and where dC stars are produced.

\acknowledgments

This research has made use of the Simbad database,
operated at CDS, Strasbourg, France.

% \clearpage

% \figcaption[]{The unusual colors of dC stars (asterisks), compared with
% disk dwarfs (filled circles), spheroid subdwarfs (x's), and white dwarfs
% (open circles). \label{fig1}}

% \figcaption[dcstarfig2.eps]{The absolute V magnitude of the three dC stars
% with measured parallaxes. \label{fig2}}

% \figcaption[dcstarfig3.eps]{The absolute K magnitude of the three
% dC stars. \label{fig3}}

% \figcaption[]{The absolute V magnitude against effective temperature
% of the one dC star with a measured temperature.  \label{fig4}}

\clearpage

\begin{deluxetable}{lcrrrrrrrrr}
\scriptsize
\tablecaption{Astrometric and Photometric Data. \label{tbl-1}}
\tablewidth{0pt}
\tablehead{
\colhead{Star} &
\colhead{R.A.(J2000){\tablenotemark{a}}} &
\colhead{$\pi_{rel}$}  &
\colhead{$\Delta \pi$} &
\colhead{$\pi_{abs}$}  &
\colhead{$\mu_{rel}$}  &
\colhead{PA}   &
\colhead{V}            &
\colhead{B$-$V} &
\colhead{K{\tablenotemark{b}}} &
\colhead{J$-$H} \\
\colhead{Names} &
\colhead{DEC.(J2000)} &
\colhead{(mas)} &
\colhead{(mas)} &
\colhead{(mas)} &
\colhead{(mas yr$^{-1}$)} &
\colhead{(deg)} &
\colhead{} &
\colhead{V$-$I} &
\colhead{} &
\colhead{H$-$K}
}
\startdata
% \vspace{-2.0 mm}
LP765$-$18 &
\phn 00 26 00.5 &
 6.88 & 1.08 & 7.96 &
613.2\phm{xxx} & 180.92 &
15.07 & 1.69 & 11.64 & 0.58 \nl
\vspace{4.0 mm}
LHS1075 &
--19 18 52\phm{xx} &
$\pm$0.80 & $\pm$0.24 & $\pm$0.84 &
$\pm$0.6\phm{xxx} & $\pm$0.03 &
& 1.61 & & 0.32 \nl
%
% \vspace{-2.mm}
LP328$-$57 &
\phn 15 52 37.4 &
 3.25 & 1.29 & 4.54 &
278.4\phm{xxx} & 228.00 &
16.30 & 1.67 & 12.82 & 0.61 \nl
\vspace{4.0 mm}
CLS96    &
+29 28 00\phm{xx} &
$\pm$0.56 & $\pm$0.35 & $\pm$0.66 &
$\pm$0.2\phm{xxx} & $\pm$0.04 &
& 1.57 & & 0.32 \nl
%
% \vspace{-2.mm}
G77$-$61{\tablenotemark{c}} &
\phn 03 32 38.0 &
15.0\phm{x} & 1.9\phm{x} & 16.9\phm{x} &
772.5\phm{xxx} & 165.80 &
13.89 & 1.75 & 10.49 & 0.64 \nl
\vspace{4.0 mm}
LHS1555 &
+01 58 00\phm{xx} &
$\pm$2.2\phm{x} & $\pm$0.4\phm{x} & $\pm$2.2\phm{x} &
$\pm$0.6\phm{xxx} & $\pm$0.10 &
& 1.56 & & 0.36 \nl
\enddata

\tablenotetext{a}{Positions are given for epoch and equinox 2000.0
   based on \cite{deu94}.}
\tablenotetext{b}{JHK photometry on the CTIO system taken from
   \cite{bot91} and \cite{dea86}.}
\tablenotetext{c}{For G77$-$61, the astrometry and BVI photometry
   are from \cite{dah77} and \cite{dea86}.}
 
\end{deluxetable}

% \clearpage

\begin{deluxetable}{lrrccrrr}
\footnotesize
\tablecaption{Derived Results. \label{tbl-2}}
\tablewidth{0pt}
\tablehead{
\colhead{Star} &
\colhead{M$_{\rm V}$} &
\colhead{M$_{\rm K}$} &
\colhead{V$_{\rm tan}$}   &
\colhead{V$_{\rm r}$} &
\colhead{U}   &
\colhead{V}  &
\colhead{W}  \\
\colhead{} &
\colhead{} &
\colhead{} &
\colhead{(km s$^{-1}$)} &
\colhead{(km s$^{-1}$)} &
\multicolumn{3}{c}{(km s$^{-1}$)}
} 
\startdata
LP765$-$18 &
 9.58 $\pm$ 0.23 &
 6.14 $\pm$ 0.23 &
365 $\pm$ 39 &
{\phn}56 $\pm$ 19 &
205 & --282 & --98  \nl
LP328$-$57 &
 9.59 $\pm$ 0.32 &
 6.11 $\pm$ 0.32 &
 291 $\pm$ 43 &
 164 $\pm$ 10    &
148 & --164 & 252  \nl
G77$-$61  &
10.03 $\pm$ 0.28 &
 6.63 $\pm$ 0.28 &
 217 $\pm$ 43 &
--34 $\pm$ {\phn}1 &
106 & --176 & --40  \nl
\enddata

\end{deluxetable}

% \end{document}

\clearpage

\begin{figure}
\plotone{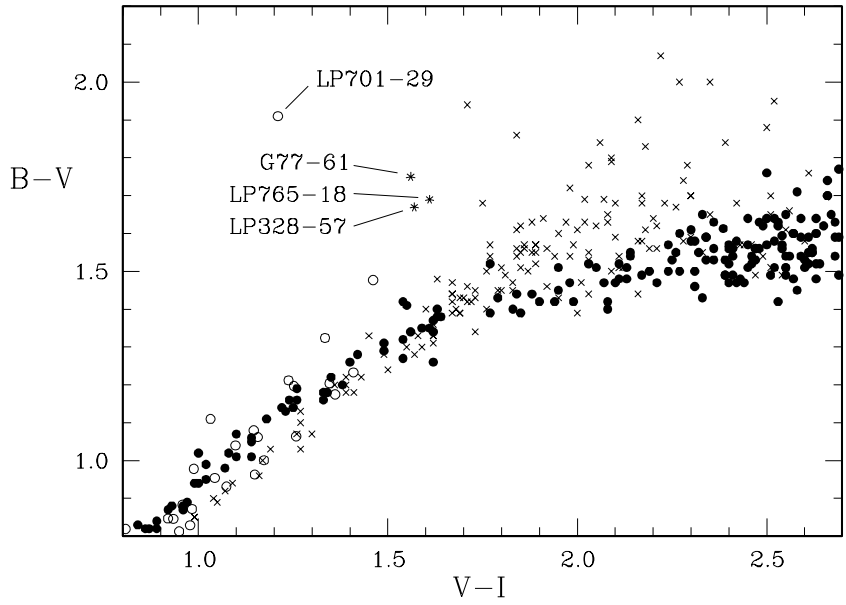}
\caption{The unusual colors of dC stars (asterisks), compared with
disk dwarfs (filled circles), spheroid subdwarfs (x's), and white dwarfs
(open circles). \label{fig1}}
\end{figure}

\begin{figure}
\plotone{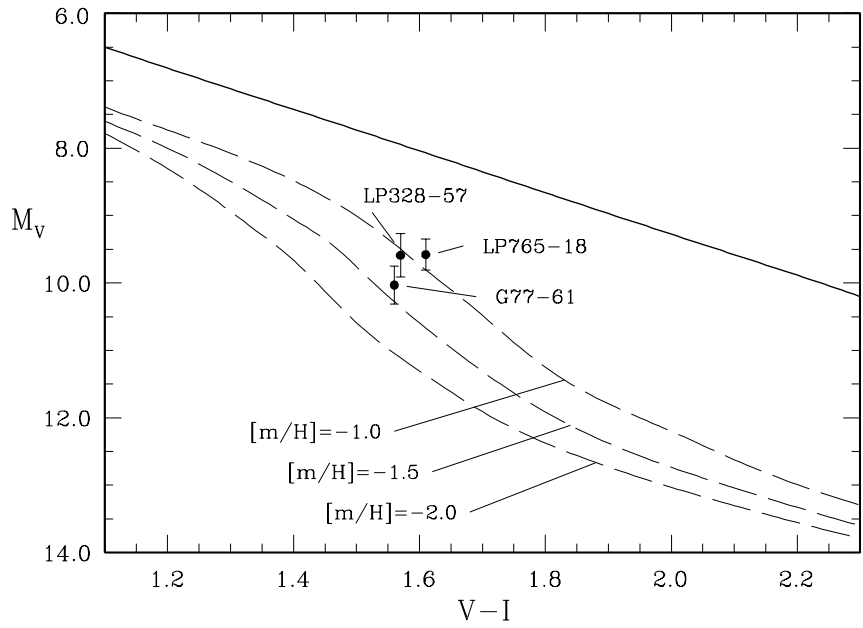}
\caption{The absolute V magnitude of the three dC stars
with measured parallaxes. \label{fig2}}
\end{figure}

\begin{figure}
\plotone{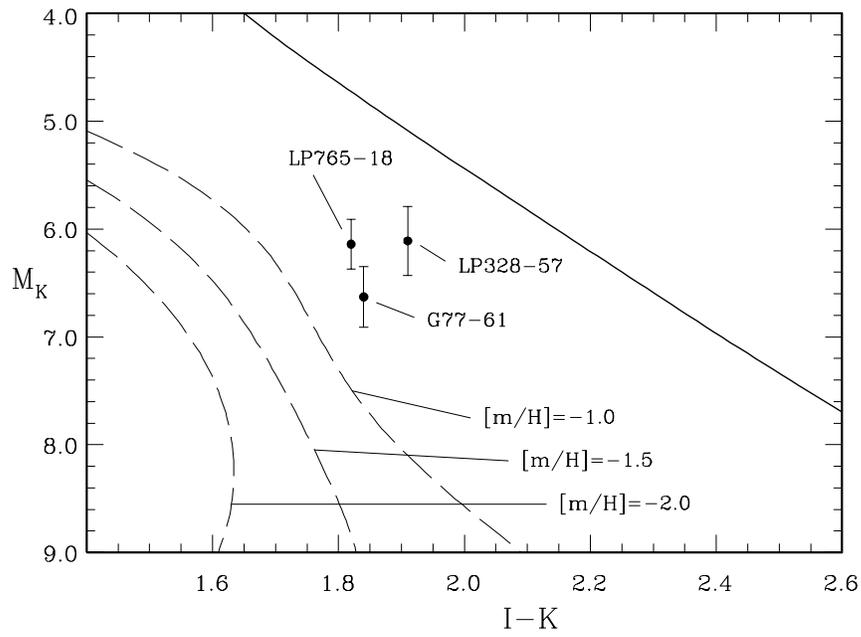}
\caption{The absolute K magnitude of the three
dC stars. \label{fig3}}
\end{figure}

\begin{figure}
\plotone{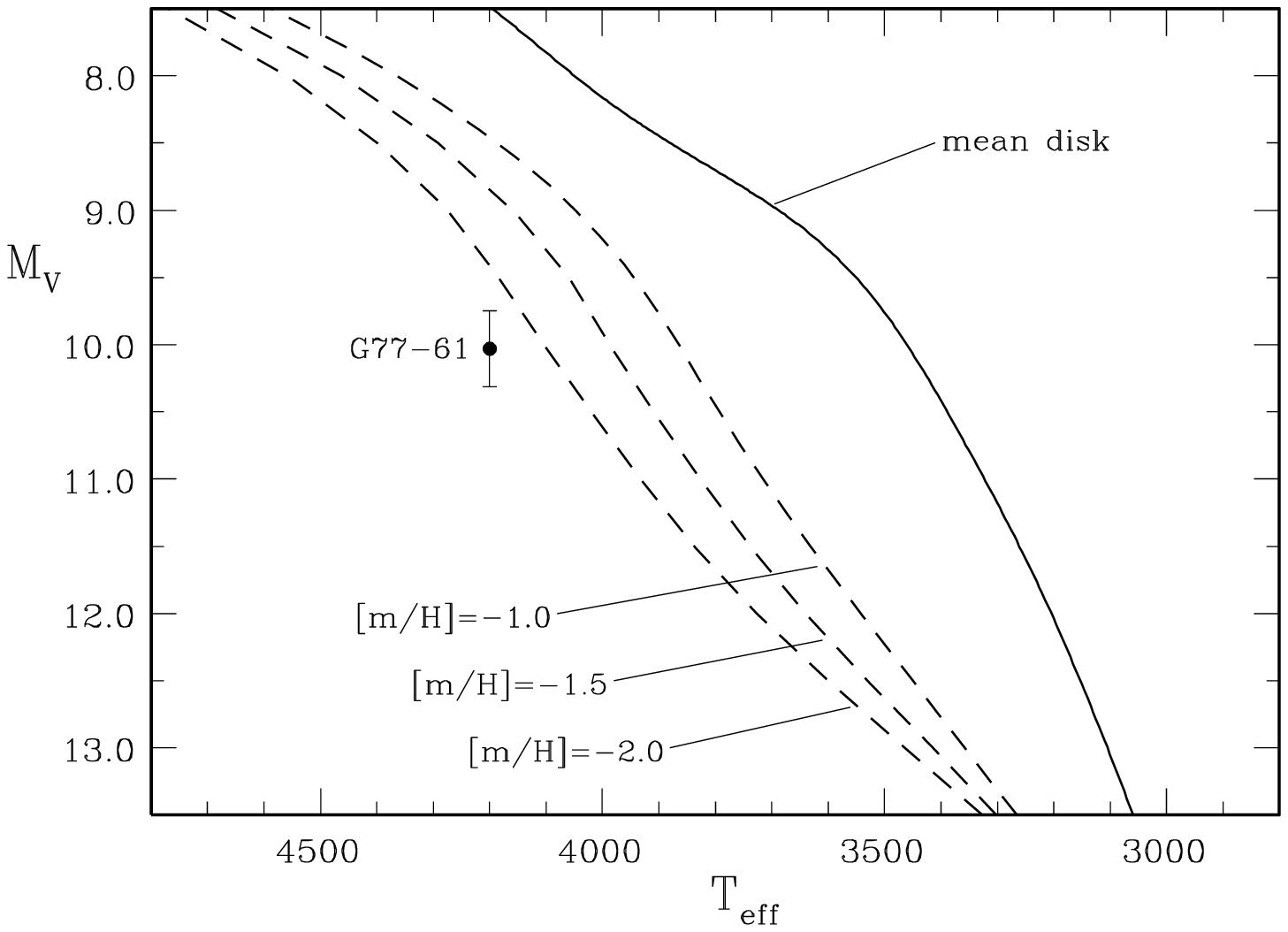}
\caption{The absolute V magnitude against effective temperature
of the one dC star with a measured temperature.  \label{fig4}}
\end{figure}


\clearpage
 
\begin{thebibliography}{}

\bibitem[Baraffe et al.\ 1997]{bar97} Baraffe, I., Chabrier, G.,
   Allard, F., \& Hauschildt, P.H. 1997, \aap, 327, 1054
\bibitem[Bothun et al.\ 1991]{bot91} Bothun, G., Elias, J.H.,
   MacAlpine, G., Matthews, K., Mould, J.R., Neugebauer, G.,
   \& Reid, I.N. 1991, \aj, 101, 2220
\bibitem[Dahn 1997]{dah97} Dahn, C.C. 1997, in Fundamental Stellar
   Properties: The Interaction between Observation and Theory,
   Proc. IAU Symp. 189, ed. T.R. Bedding, A.J. Boothe, \& J. Davis
   (Dordrecht: Kluwer Academic), 19
\bibitem[Dahn et al.\ 1977]{dah77} Dahn, C.C., Liebert, J., Kron, R.G.,
   Spinrad, H., \& Hintzen, P. 1977, \apj, 216, 757
\bibitem[Dahn et al.\ 1978]{dah78} Dahn, C.C., Hintzen, P.M., Liebert,
   J.W., Stockman, H.S., \& Spinrad, H. 1978, \apj, 219, 979
\bibitem[Dearborn et al.\ 1986]{dea86} Dearborn, D.S.P., Liebert, J.,
   Aaronson, M., Dahn, C.C., Harrington, R., Mould, J.,
   \& Greenstein, J.L. 1986, \apj, 300, 314
\bibitem[de Kool \& Green 1995]{dek95} de Kool, M., \& Green, P.J.
   1995, \apj, 449, 236
\bibitem[Deutsch 1994]{deu94} Deutsch, E.W. 1994, PASP, 106, 1134
\bibitem[Gass 1988]{GAS88} Gass, H. 1988, \aap, 193, 185
\bibitem[Gass et al.\ 1988]{gas88} Gass, H., Liebert, J., \&
   Wehrse, R. 1988, \aap, 189, 194
\bibitem[Green 1997]{gre97} Green, P.J. 1997, in The Carbon Star
   Phenomenon, Proc. IAU Symp. 177, ed. xxx (Dordrecht: Kluwer), in press
\bibitem[Green et al.\ 1992]{gre92} Green, P.J., Margon, B.,
   Anderson, S.F., \& MacConnell, D.J. 1992, \apj, 400, 659
\bibitem[Leggett 1992]{leg92} Leggett, S.K. 1992, \apjs, 82, 351
\bibitem[Monet et al.\ 1992]{mon92} Monet, D.G., Dahn, C.C.,
   Vrba, F.J., Harris, H.C., Pier, J.R., Luginbuhl, C.B., \&
   Ables, H.D. 1992, \aj, 103, 638
\bibitem[Proffitt \& Michaud 1989]{pro89} Proffitt, C.R.,
  \& Michaud, G. 1989, \apj, 345, 998
\bibitem[van Altena et al.\ 1995]{van95} van Altena, W.F., Lee,
   J.T., \& Hoffleit, E.D. 1995, The General Catalogue of Trigonometric
   Stellar Parallaxes, Fourth Edition (L. Davis Press: Schenectady)

\end{thebibliography}
\end{document}